\documentstyle[aps,epsfig]{revtex}

\title{Synchronization  under  periodic  modulation  of potential
wells in a two-state stochastic system}

\author{Asish K. Dhara and S. R. Banerjee}
\address{Variable Energy Cyclotron  Centre,
1/AF, Bidhan  Nagar,  Calcutta-700064,India}

\begin{document}
\maketitle
\date{today}

\begin{abstract}
We  analyse  the  effect  of  synchronization  between  noise and
periodic  signal  in  a  two-state  spatially   extended   system
analytically.  Resonance  features  are demonstrated. To have the
maximum cooperation between signal and noise, it  is  shown  that
noise  strength  at  resonance  should increase linearly with the
frequency of the  signal.  The  time  scale  of  the  process  at
resonance  is  also shown to increase linearly with the period of
the signal.

\end{abstract}


Noise  induced  large  response  in  bistable potential to a weak
periodic signal commonly referred to as stochastic resonance (SR)
attracts   considerable    interests    during    recent    years
\cite{Mos,Gam1,Gri}. Manifestation of SR is usually exihibited as
non-monotonous  behavior of power spectrum of the process or as a
similar  behavior  of  the  peak  heights  of   the   escape-time
distribution  related  to  switching  between  two  wells,  as  a
function of the noise strength \cite{Mac,Fox1,Jun,Zho}. The usage
of the term "resonance" is then criticised  based  on  two  facts
\cite{Fox2}.  The  value  of  the  noise  strength  for which the
maximum of the spectral density at forcing frequency occurs  does
not  simply  correspond to the period of the signal and for fixed
noise strength, the spectral density decreases monotonically with
frequency of the signal, at variance with the notion of  ordinary
resonance.  One  therefore would naturally be curious whether any
characterisation  of  this  phenomena  in  terms  of  our   usual
definition of resonance is possible as it was originally pictured
\cite{Ben}.

Recently,  it  has  been  found  that  such description is indeed
possible  \cite{Gam2}.  Simulating  the   continuous   stochastic
process  into  a  stochastic  point  process  with the help of an
analog circuit they \cite{Gam2} measure the strength of the first
few peaks of the residence time distribution. It  is  found  that
when  strength  of the first peak is plotted as a function of the
frequency of the signal keeping noise strength constant, it  hits
a  maximum for that period which is equal to twice of the inverse
of the Kramers rate for the unbiased process for that  particular
noise  strength.  Alternately, when strength of the first peak is
plotted as a function of noise strength it attains a  maximum  at
that  value  of  the  noise strength for which inverse of Kramers
rate for the unbiased process is equal to the period  of  forcing
signal.    Subsequently,   this   resonant   behavior   is   also
experimentally verified in the  polarised  emission  of  vertical
cavity  surface  emitting  laser  \cite{Gia}.  In  this  paper we
analyse these results analytically.

\vspace{0.75cm}
\begin{figure}[htb]
\centerline{\psfig{file=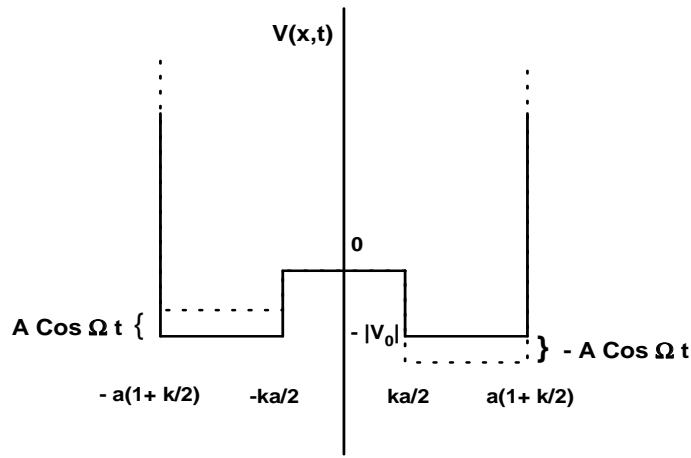, height=6cm, width=9cm}}
\caption{Potential $V(x,t)$ at time $t$ as a function of $x$. }
\label{fig.1}
\end{figure}

Response  of  a two-state spatially extended system embedded to a
noisy environment under the influence of a periodic field  is  of
our  concern.  The  simplest  model  \cite{Ber} for the two-state
spatially extended system is to consider a particle moving in the
piecewise double well potential $V(x)$  under  the  influence  of
white  noise.  The  potential  $V(x)$  is  shown  in  Fig.1.  The
influence of the external periodic field is usually described  by
the  modulation  of  the potential well in the following fashion.
The potential at any instant of time $t$ for the  left  state  is
replaced  by  $-\mid  V_0 \mid + A cos\Omega t $ and that for the
right state is replaced by $-\mid V_0 \mid - A cos\Omega t $ with
$ A,\Omega $ being the amplitude and frequency  of  the  periodic
signal and $\mid V_0 \mid$ is the barrier height when there is no
modulation.

The Fokker-Planck equation (FPE) for the probability distribution
function  $P(x,t)$  for  position $x$ of the particle at time $t$
for this process is

\begin{equation}
\label{Eq.1}
\frac{\partial P(x,t)}{\partial t} = \frac{\partial}{\partial x}
\left[\frac{\partial V(x,t)}{\partial x}P(x,t)\right]
+ D \frac{\partial^2P(x,t)}{\partial x^2}~,
\end{equation}
where  $D$  is  strength of the white noise. It is clear that for
the  potential  in  Fig.1,  $\frac{\partial  V}{\partial  x}  =0$
everywhere  except at the discontinuous points. Therefore in each
region of $V(x, t)$ =constant, the  FPE  (1)  reduces  to  simple
diffusion equation. The solution in each region are to be matched
with the continuity of probability current and jump conditions at
the  discontinuous  points  [$-ka/2,  ka/2$]  at each time. These
conditions are to be  supplemented  by  the  reflecting  boundary
conditions at the walls [$-a(1+k/2), a(1+k/2)$].

In  this  paper  we  however are interested in first passage time
density function (FPTDF) with  mean  first  passage  time  (MFPT)
being  a  important  parameter  of the process. To be specific we
concentrate on the events starting from $ x=-1$ and ending  at  $
x=1$.  As  the potential is symmetric, the MFPT $<t(-1\rightarrow
1)> $  would  be  same  as  $<t(1\rightarrow  -1)>  $.  Thus  the
conditions that the required solution of Eq.(1) satisfies are

\begin{mathletters}
\label{Eq.2}
\begin{eqnarray}
P'(-a(1+k/2), t) &=& 0~, \\
e^{[-\mid V_0 \mid + A cos\Omega t ]/D} P(-ka/2-0, t) &=& P(-ka/2+0, t)~, \\
P'(-ka/2-0, t) &=& P'(-ka/2+0, t)~, \\
P(ka/2-0, t) &=& e^{[-\mid V_0 \mid - A cos\Omega t ]/D} P(ka/2+0, t)~, \\
P'(ka/2-0, t) &=& P'(ka/2+0, t)~, \\
P(1, t) &=& 0~,
\end{eqnarray}
\end{mathletters}
where  prime  over  $P$  in the above equations denote derivative
with respect to $x$. Eq.(2a) corresponds to  reflecting  boundary
condition  for  the wall at $ x = -a(1+k/2)$, Eq.(2b) and Eq.(2d)
are the jump conditions at the discontinuous points $ x= \pm ka/2
$, Eq.(2c) and Eq.(2e) are the matching condition for  continuity
of  probability  current  and  Eq.(2f)  is the absorbing boundary
condition ar $ x=1$ for the process considered.

For $(A/D)$ small, we expand the solution in each region as

\begin{equation}
\label{Eq.3}
P(x,t) = e^{[-\frac{\mid V_0 \mid}{ 2D}-\lambda t - (A/D) g(t)]}
\psi (x) [1+\sum_{m=1}^{\infty}(A/D)^m f_m(x, t)] ~,
\end{equation}
with $f_0 = 1$, and $g(t)$ being defined as
\begin{eqnarray}
\label{Eq.4}
g(t)=&& cos \Omega t~ ; -a(1+k/2)\leq x \leq-ka/2 ~, \nonumber \\
=&& 0~ ;       -ka/2 < x < ka/2 ~,\\
=&&-cos \Omega t~ ;ka/2 \leq x \leq 1~.\nonumber
\end{eqnarray}
In  Eq.(3),  $\lambda$  is  some  constant  and  $f_m(x,  t)$ are
functions, which are to be determined.  Substituting  the  ansatz
(3)  in Eq.(1) and equating equal powers of $(A/D)$ we obtain the
following set of equations in each region:

\begin{mathletters}
\label{Eq.5}
\begin{eqnarray}
\frac{\partial^2\psi(x)}{\partial x^2} &=&  -(\lambda/D)
\psi(x)~, \\
\left[\frac{ \partial}{ \partial t}
- D \frac{ \partial^2 }{ \partial x^2}\right]
[\psi (x) f_n(x, t)] &=& \dot{g}(t) f_{n-1}(x, t)~; n =1,2,...~,
\end{eqnarray}
\end{mathletters}
with  $f_0  =  \psi  (x)$. For small $(A/D)$ we keep only Eq.(5a)
which is obtained after equating the terms O(1). Substituting the
approximate form  of  probability  density  in  Eqs.(2a)-(2f)  we
obtain the following set of matching conditions for the functions
$\psi (x)$.

\begin{mathletters}
\label{Eq.6}
\begin{eqnarray}
\psi_{\mu}'(-a(1+k/2)) &=& 0~, \\
e^{[-\mid V_0 \mid ]/D} \psi_{\mu}(-ka/2-0) &=&
\psi_{\mu}(-ka/2+0)~, \\
\psi_{\mu}'(-ka/2-0) &=&\psi_{\mu}'(-ka/2+0)~, \\
\psi_{\mu}(ka/2-0) &=& e^{[-\mid V_0 \mid ]/D}
\psi_{\mu}(ka/2+0)~, \\
\psi_{\mu}'(ka/2-0) &=& \psi_{\mu}'(ka/2+0)~, \\
\psi_{\mu}(1) &=& 0~,
\end{eqnarray}
\end{mathletters}
We  note  that  the  solutions  of  Eqs.(6a)-(6f)  are  the exact
eigenfunctions of  the  corresponding  unperturbed  process  with
associated    eigenvalues   $\lambda_{\mu}$.   Thus   with   this
approximation  the  conditional  probability  for   getting   the
particle  at  position  $x$ at time $t$ when it is known to start
from $ x=-1$ at $ t=0 $ is

\begin{equation}
\label{Eq.7}
P(x, t\mid -1, 0) = e^{-[V(x)-V(-1)]/2 D} e^{-A[g(t)-1]/D}
\sum_n e^{-\lambda_n t}\psi_n(x)\psi_n(-1)~,
\end{equation}
where  the  value  of  $V(x)$  and  $g(t)$  would be taken as the
respective values of the region where $x$  falls.  As  ${\psi_n}$
form  the complete set, $ P(x, t=0+\mid -1, 0) = \delta (x+1)$ is
automatically satisfied. The Eqs. (6a)-(6f) are readily solved to
obtain  the  full  set  of   eigenfunctions   and   corresponding
eigenvalues. They express as

\begin{mathletters}
\label{Eq.8}
\begin{eqnarray}
\psi_{\mu}(x)=&& C_{\mu} cos \{k_{\mu}[x+a(1+k/2)]\}~;
-a(1+k/2)\leq x <-ka/2~,\\
=&& -C_{\mu}[e^{\mid V_0 \mid/2D} sin (k_{\mu}a) sin \{k_{\mu}
(x+ka/2)\}\nonumber\\
&&-e^{-\mid V_0 \mid/2D} cos (k_{\mu}a)
cos \{k_{\mu}(x+ka/2)\}]~; -ka/2< x <ka/2 ~,\\
=&&C_{\mu}[ cos(k_{\mu}ka) cos \{k_{\mu}[x+a(1-k/2)]\}\nonumber\\
&&-e^{\mid V_0 \mid/D} sin (k_{\mu}ka) sin (k_{\mu}a)
cos \{k_{\mu}(x-ka/2)\}\nonumber\\
&&-e^{-\mid V_0 \mid/D} sin (k_{\mu}ka) cos (k_{\mu}a)
sin \{k_{\mu}(x-ka/2)\}]\nonumber\\
&&~; ka/2< x \leq 1~,
\end{eqnarray}
\end{mathletters}
where  constant  $C_{\mu}$  is  determined from the normalisation
condition

\begin{equation}
\label{Eq.9}
C_{\mu}^2\int_{-1}^1\psi_{\mu}^2(x) dx = 1~.
\end{equation}
The   corresponding   eigenvalues   are   determined   from   the
transcendental equation

\begin{eqnarray}
\label{Eq.10}
cos(k_{\mu}ka) cos \{k_{\mu}[1+a(1-k/2)]\}\nonumber\\
-e^{\mid V_0 \mid/D} sin (k_{\mu}ka) sin (k_{\mu}a)
cos \{k_{\mu}(1-ka/2)\}\nonumber\\
-e^{-\mid V_0 \mid/D} sin (k_{\mu}ka) cos (k_{\mu}a)
sin \{k_{\mu}(1-ka/2)\}=0~.
\end{eqnarray}
The  expressions  (8)  and  $k_{\mu}$  obtained  from Eq.(10) are
substituted in Eq.(7)  with  $\lambda_{\mu}  =  D  k_{\mu}^2$  to
obtain  the  conditional  probability  $P(x, t\mid -1, 0)$ at any
point $x$ at time $t$.

The  survival  probability  at  time $t$ , $S(t)$ can be obtained
from Eq.(7) as

\begin{equation}
\label{Eq.11}
S(t) = \int_{-1}^1 P(x, t\mid -1, 0) dx ~,
\end{equation}
and  the  first  passage  time density function (FPTDF) $g(t)$ is
obtained from $S(t)$ as

\begin{equation}
\label{Eq.12}
g(t) = - \frac{d S(t)}{dt}~.
\end{equation}
Physically,  $g(t)  dt$  gives  the probability that the particle
reaches $x=1$ for the first time in the  time  interval  $t$  and
$t+dt$  starting  from  $x=-1$ at $t=0$. As the expression of the
conditional probability (7) is approximate and strictly valid for
small $(A/D)$, we should demand the necessary physical conditions
like positivity, normalisability on $g(t)$. Different moments  of
$g(t)$ can be readily calculated from the normalised $g(t)$ as
\begin{equation}
\label{Eq.13}
<t^j> = \int_0^{\infty}  t^j g(t)dt ~; j=1,2,...~.
\end{equation}

\vspace{0.5cm}
\begin{figure}[htb]
\begin{center}
\begin{tabular}{lr}
\psfig{file=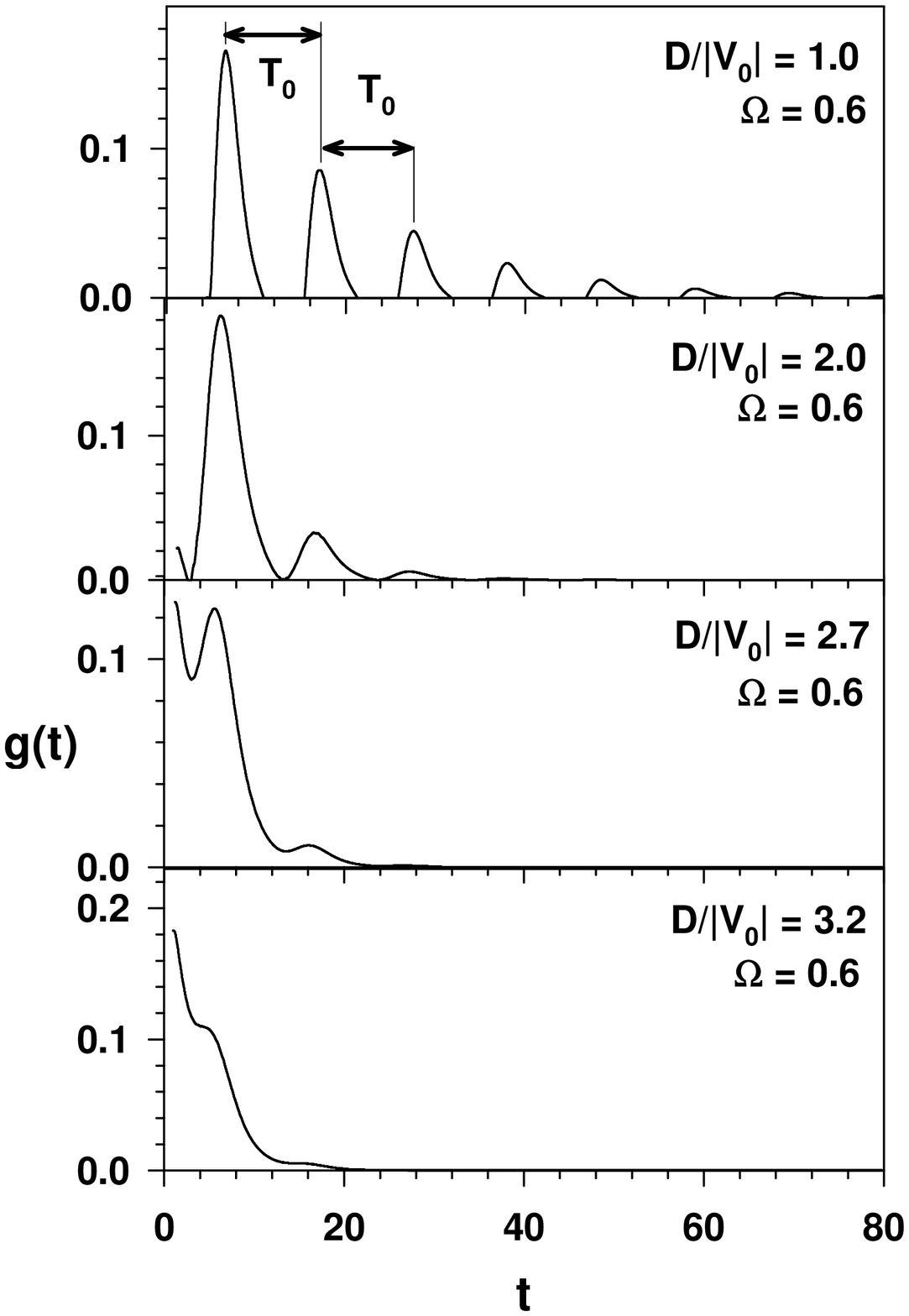, height=6cm, width=6cm} \hspace{1cm}
\psfig{file=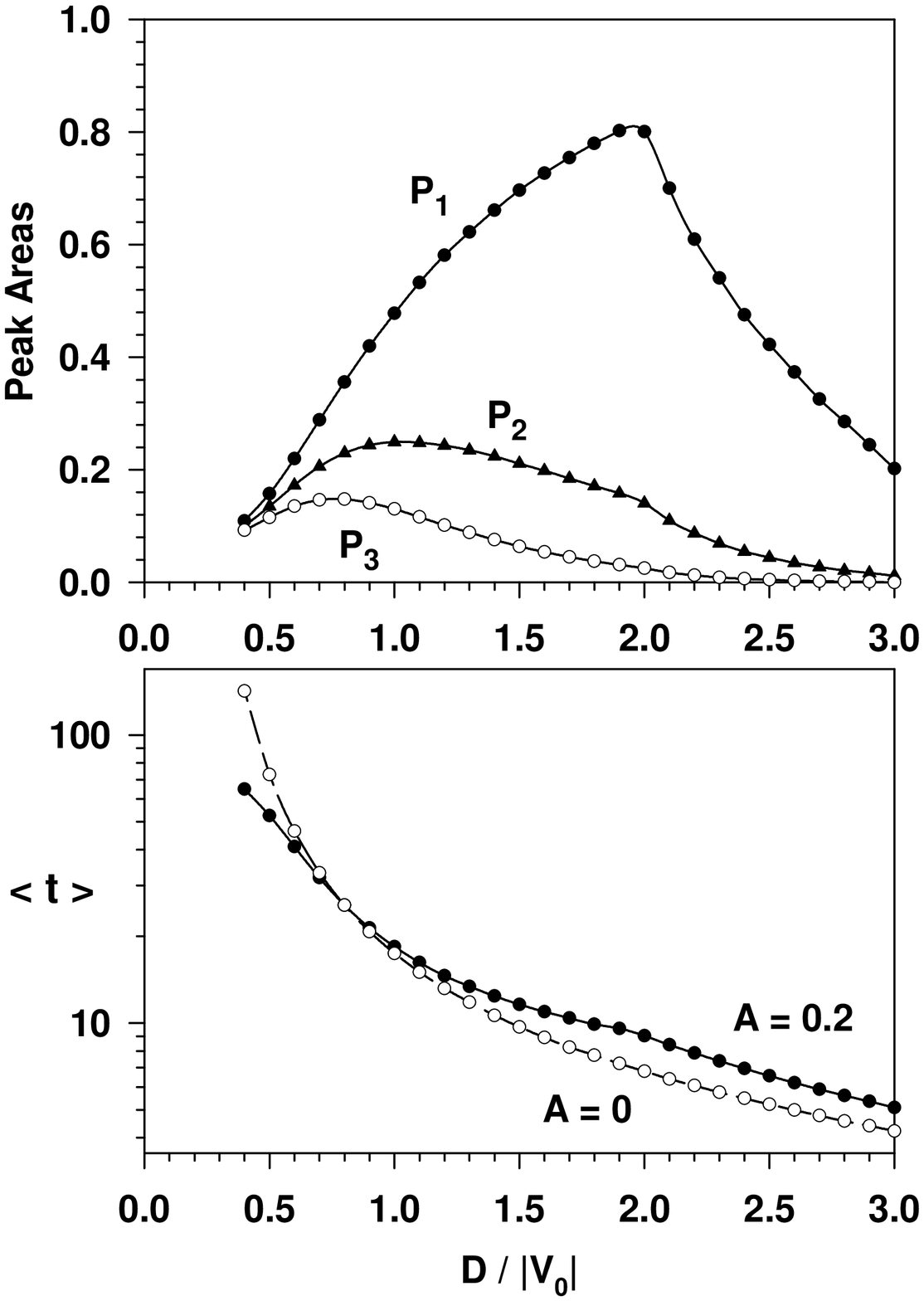, height=6cm, width=6cm}
\end{tabular}
\begin{minipage}[t]{3in}
\caption{FPTDF  $g(t)$ as a function of $t$ for
frequency $\Omega$ = 0.6 and $(D/V_0)$ = 1.0, 2.0, 2.7, 3.2.}
\end{minipage} \ \hspace{0.5cm} \
\begin{minipage}[t]{3in}
\caption{Peak areas as a function of $ (D/V_0)$
and MFPT $<t>$ as
a  function  of $(D/V_0)$ with signal $(A=0.2)$ [line with filled
circles] and without signal $(A = 0)$ [line with open circles].}
\end{minipage}
\end{center}
\end{figure}

The  calculations  are  done with the potential parameters, $a=1,
k=1, \mid V_0 \mid = 0.25, A=0.2 $. Normalised FPTDF $g(t)$ as  a
function of $t$ for fixed value of signal frequency $\Omega = 0.6
$  and  different  values  of  $(D/\mid V_0 \mid)$ are plotted in
Fig.2. Multiple peaks are observed  for  small  values  of  noise
strength  and  as  noise  strength  increases background is found
dominant. $g(t)$ extends  for  a  longer  time  for  small  noise
strength. The peak heights fall exponentially with time for fixed
$(D/\mid  V_0  \mid)$  but  peak  heights  are increased as noise
strength is lowered. Successive peaks are  separated  by  a  time
exactly  equal to the period of the signal frequency showing that
the probability of transition is maximum after  each  period.  At
these particular times coperation between noise and the signal in
making  a  transition  is  more. It clearly demonstrates that the
signal synchronizes the  noise  induced  transition.  Peak  areas
(counting  peaks from left ) are also found to be decreasing with
time.

On  subtracting the exponential background, the peak areas $P_n =
\int g(t) dt $,where the integration is done around $n$ th.  peak
are  plotted  as  a function of $(D/\mid V_0 \mid)$ in Fig.3. The
figure demonstrates  a  non-monotonous  behavior  for  each  peak
exihibiting  a signature of maximum coperation or synchronization
between noise and external signal  at  specific  value  of  noise
strength for a particular signal frequency $\Omega$. The position
of  the  maximum  of  $P_n$  shifts  towards lower value of noise
strength as peak number increases. All curves merges to  low  $D$
showing  at  very  low  noise strength transition would mainly be
controlled by signal most of the time.

\vspace{1cm}
\begin{figure}[htb]
\begin{center}
\begin{tabular}{lr}
\psfig{file=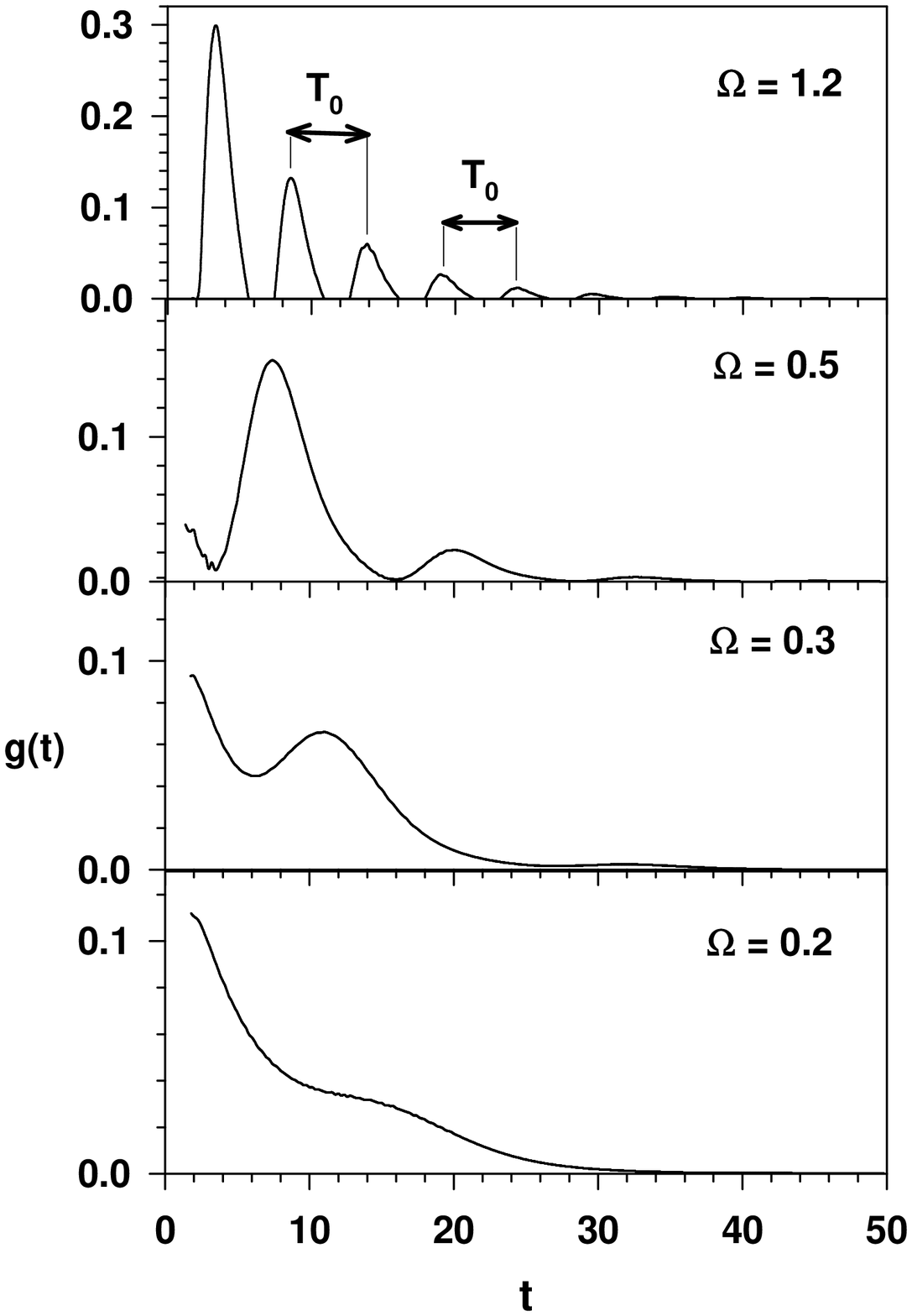, height=6cm, width=6cm} \hspace{1cm}
\psfig{file=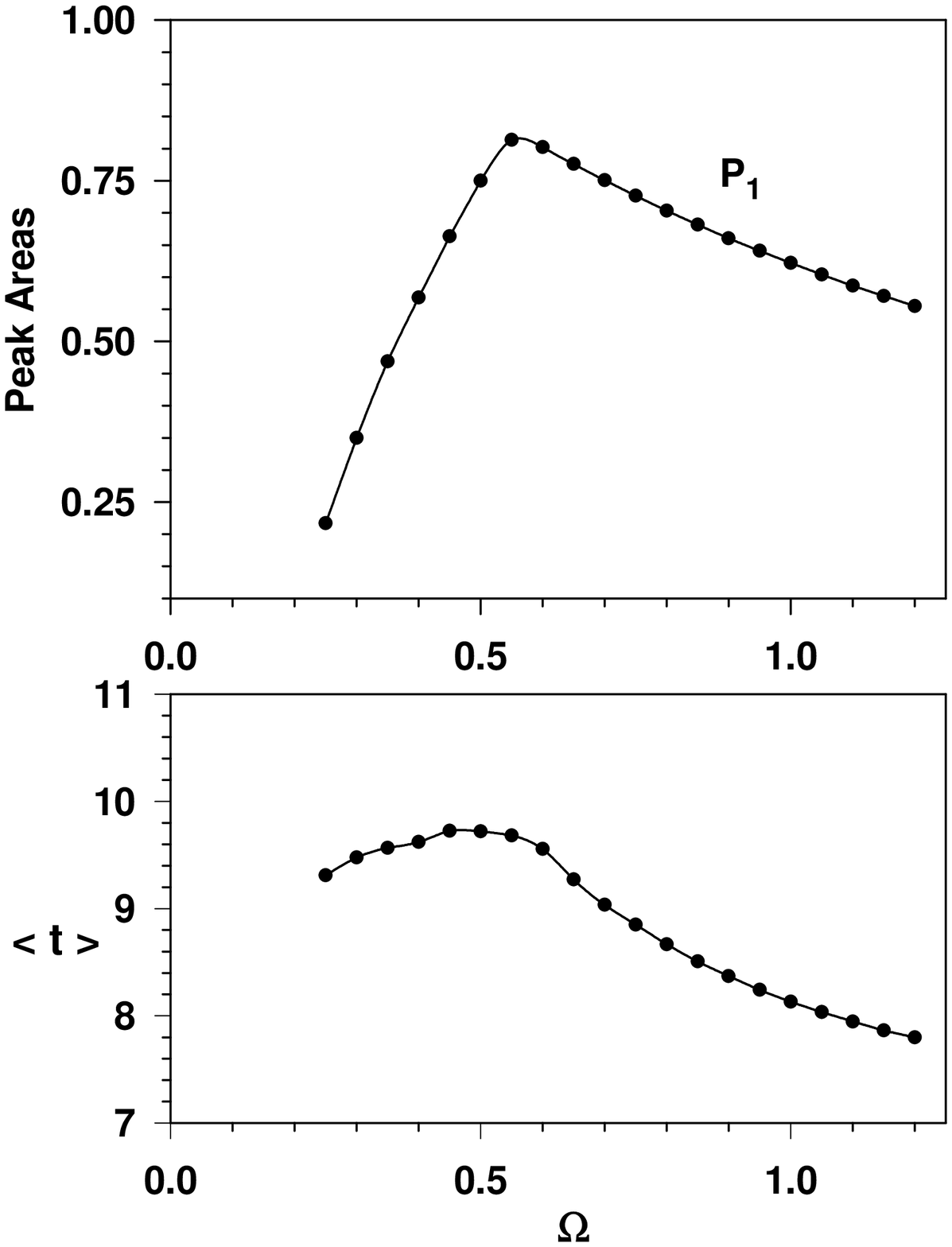, height=6cm, width=6cm}
\end{tabular}
\begin{minipage}[t]{3in}
\caption{FPTDF  $g(t)$  as function of time $t$ for $D/V_0 = 1.9$
and $\Omega = 0.2, 0.3, 0.5, 1.2. $}
\end{minipage} \ \hspace{0.5cm} \
\begin{minipage}[t]{3in}
\caption{Peak areas as a function of $\Omega$ and MFPT $<t>$ as a
function of $\Omega$. }
\end{minipage}
\end{center}
\end{figure}

In  a  similar manner, for a fixed $(D/\mid V_0 \mid =1.9 )$, the
normalised  FPTDF  $g(t)$  are  plotted  for   different   signal
frequencies  in  Fig.4.  It  also  shows  peak structure over the
background. As frequency increases the peaks  are  more  dominant
and  for  a  given frequency peak heights fall exponentially with
time. Peaks are separated by the  period  $T_0$  of  the  signal.
After  subtracting  the  exponential background, the peak area of
the first peak (which is the  most  dominant)  is  plotted  as  a
function  of  frequency of the signal in Fig.5. It also exihibits
the  non-monotonous  behaviour.  The   curve   shows   that   the
synchronization  (defining the strength of the synchronization as
the area of the peak) is maximum at specific value of the  signal
frequency.

The  non-monotonous  behaviour of the peak strength as a function
of  noise  strength  or  frequency  clearly  indicates  that  the
cooperation  between noise and signal becomes maximum at specific
value of noise strength for given frequency or at specific  value
of  signal  frequency  for  given  noise  strength.  This  is the
resonant behaviour in accordance with our usual convention.  Thus
this  specific  value  of noise strength and signal frequency are
denoted as $D_{res}, \Omega_{res}$ respectively.

As  MFPT  $<t>$  is  a  very  important  time scale in this noise
induced transition, we also plot $<t>$ after evaluating them from
Eq.(13) as a function of $(D/\mid V_0 \mid)$ in Fig.3  and  as  a
function  of  signal frequency $\Omega$ in Fig.5. For the sake of
comparison MFPT when there is  no  signal  $(A  =  0)$,  is  also
plotted as a function of $(D/\mid V_0 \mid)$ in Fig.3. The curves
show  a  monotonous fall as noise strength or frequency increases
except a little curb at low frequency.

\vspace{0.5cm}
\begin{figure}[htb]
\begin{center}
\begin{tabular}{lcr}
\psfig{file=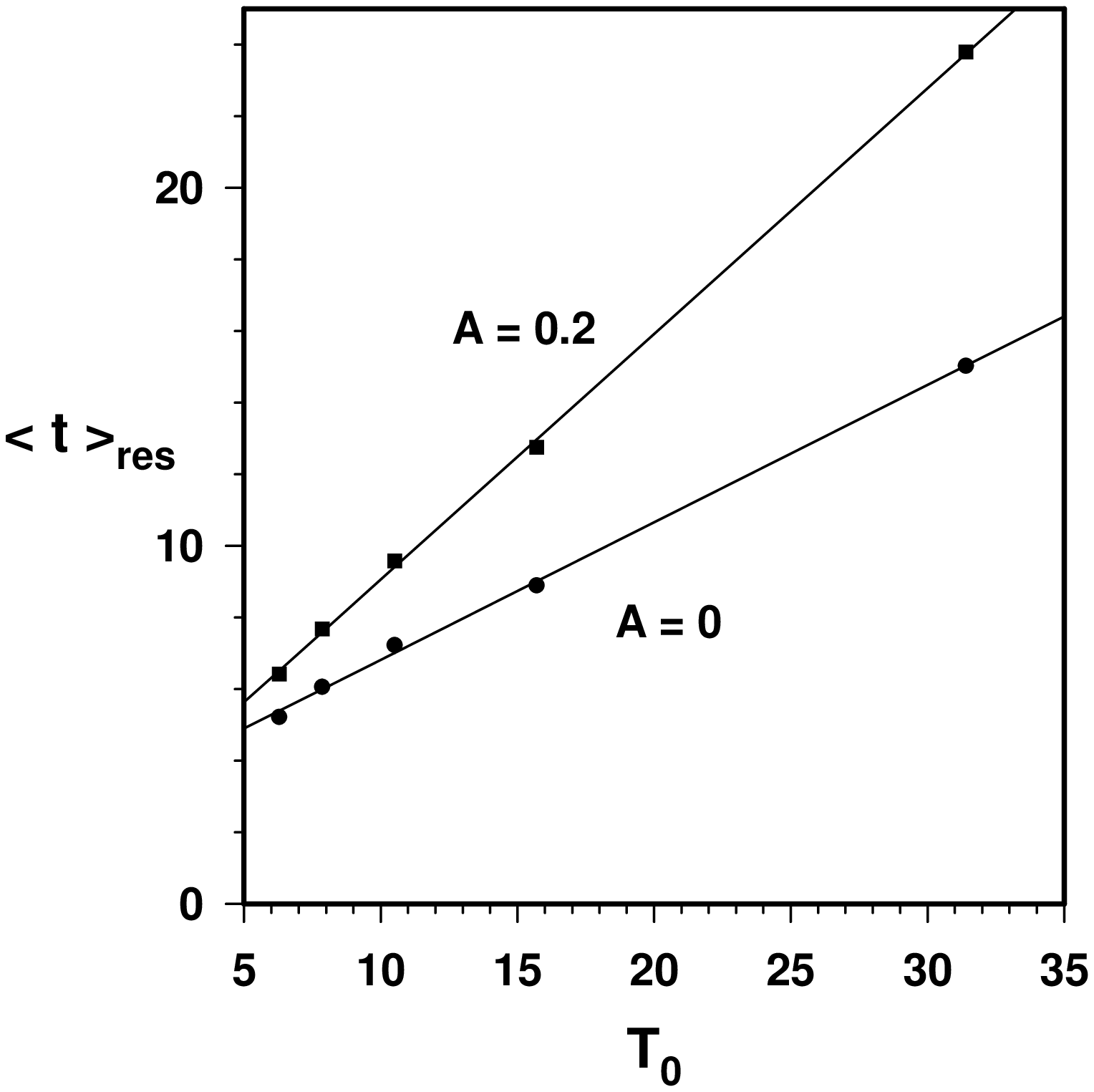, height=6cm, width=4.5cm} \hspace{1cm} 
\psfig{file=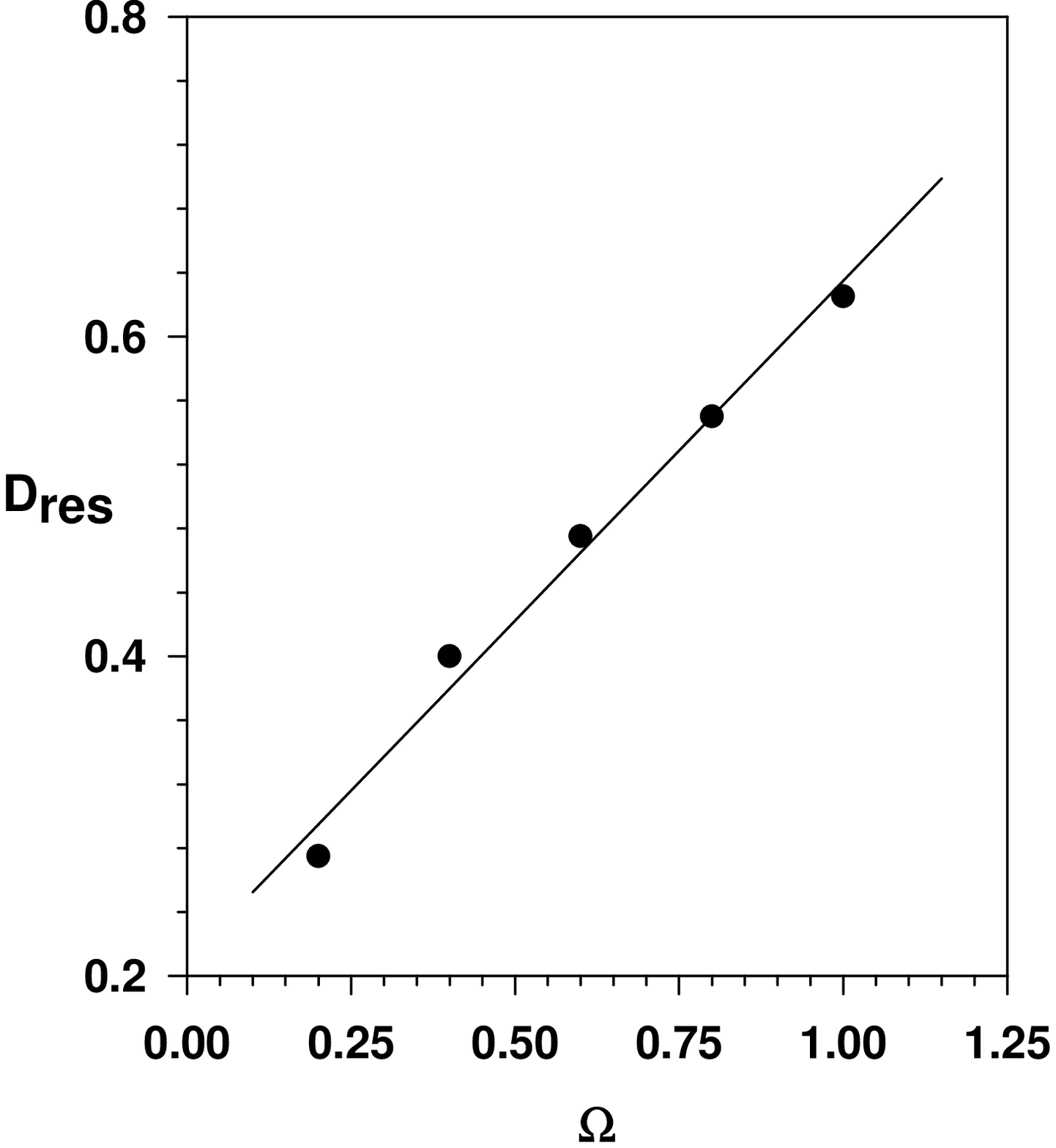, height=6cm, width=4.5cm} \hspace{0.5cm} 
\psfig{file=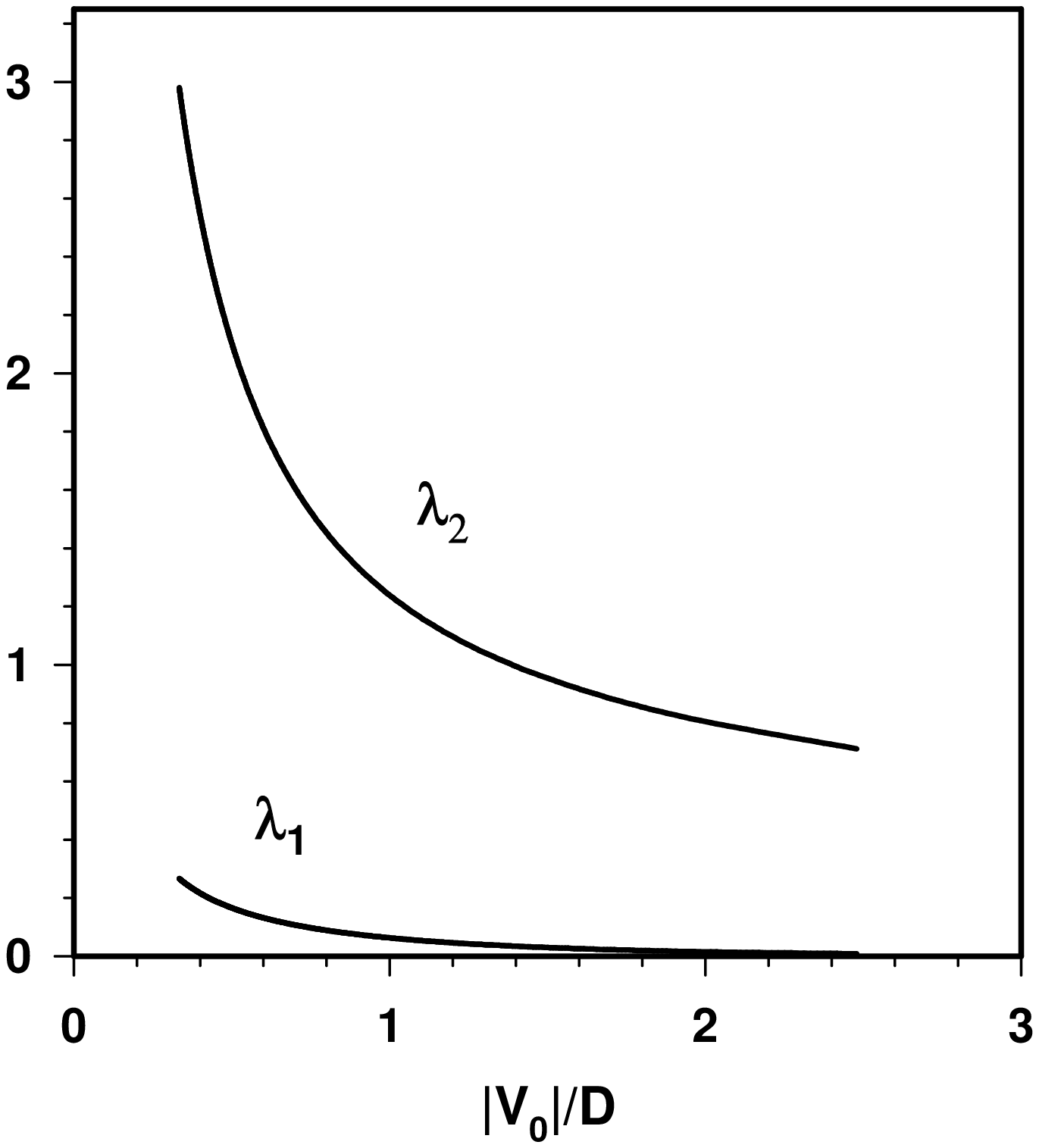, height=6cm, width=4cm}
\end{tabular}
\begin{minipage}[t]{2in}
\caption{MFPT at resonance $<t>_{res}$ as a function of period of
the  signal  $  T_0$  with  signal  $(A = 0.2)$ [line with filled
squares]  and  without  signal  $(A  =  0)$  [line  with   filled
circles].}
\end{minipage} \ \hspace{0.25cm} \
\begin{minipage}[t]{2in}
\caption{Noise strength at resonance $D_{res}$ as a function of
frequency $\Omega$}
\end{minipage} \ \hspace{0.25cm} \
\begin{minipage}[t]{1.75in}
\caption{Eigenvalues   $\lambda_{1}$   and   $\lambda_{2}$  as  a
function of $(\mid V_0\mid/D) $}
\end{minipage}
\end{center}
\end{figure}

From  the  resonance  values  $(D_{res}/\mid  V_0 \mid)$ (for the
first dominant peak) for different  frequencies  $\Omega  =  0.2,
0.4, 0.6, 0.8, 1.0 $, we evaluate MFPT $<t>_{res}$ from Fig.3 and
plot  as  a  function of period $T_0$ of the signal in Fig.6. The
points fit a straight line :$<t>=.686 T_0 +  2.2$.  In  order  to
have   a   comparison   with   MFPT  for  these  resonant  values
$(D_{res}/\mid V_0 \mid)$ for non-biased process  $(A=0)$,  these
values  are also plotted as a function of the period $T_0$ of the
signal in the same Fig.6. They also fit  straightline  :$<t>=.384
T_0  +2.985$.  This  figure  shows  that  as period of the signal
increases the resonant time scale increases exactly in  a  linear
fashion.

We  have  seen that synchronization is maximum for specific noise
strengths $D_{res}$ for different  signal  frequencies  $\Omega$.
With these resonant values, $D_{res}$ is plotted as a function of
signal  frequency  $\Omega$  in  Fig.7.  The curve is fitted to a
straight line $[D_{res} = .425\Omega + .21]$, which shows that as
frequency increases the noise strength for which resonance occurs
also should increase linearly.

In  conclusion,  we  analyse  the  stochastic  synchronization in
two-state  spatially  extended  system  under  the  influence  of
periodic  field  analytically.  The  analytic  expression  of the
conditional probability could very well  exihibit  the  resonance
behaviour  and synchronization between noise and periodic signal.
The expression can be further approximated with  only  the  first
term  in the summation by noting that the second and consequently
second onward eigenvalues (as the eigenvalues are ordered) differ
from it by a large amount. The eigenvalues $\lambda_1,  \lambda_2
$  are  plotted as a function of $(\mid V_0\mid/D)$ in Fig.8. The
figure clearly shows that for a very large variation of the noise
strength this approximation is fairly accurate.

In  asymptotic  expansion as given in Eq.(3) we consider only the
first term. From Eq.(5b) one can estimate the magnitude of  $f_1$
by   expanding   $\psi_r(x)f_1^{(r)}$  in  the  complete  set  of
eigenfunctions $\{\psi_n(x)\}$ at each time $t$. It can be  shown
that  $f_1$  would  be  a  periodic  function  and independent of
position $x$. In fact, $f_1^{(r)}(t)\sim \Omega Im[e^{i\Omega t}/
(\lambda_r+i\Omega)] $ and consequently $\mid f_1^{(r)}(t)\mid  <
\frac{1}  {[1+(\lambda_r  /\Omega)^2]^{1/2}} < 1$. Thus for small
$(A/D)$ the approximation (7) is fairly good.

The  process  (1) has in built scaling properties. When we change
time as $t\rightarrow Dt$, $\mid V_0\mid$ would change to  $(\mid
V_0\mid/D)$, $\Omega\rightarrow (\Omega/D)$, $A\rightarrow (A/D)$
respectively.

The  MFPT  of  the  process  at  resonance  is  shown to increase
linearly with the  period  of  the  signal.  The  resonant  noise
strength is also shown to increase linearly with frequency of the
signal. This fact might help the experimentalists in studying the
resonance behaviour in complex systems where tuning the frequency
is   a  convenient  task  than  varying  the  internal  noise  or
temperature.


\end{document}